\documentclass[11pt,a4paper,english,nofootinbib,,superscriptaddress]{revtex4}
\usepackage{lmodern}
\usepackage{slashed}

\usepackage[T1]{fontenc}
\usepackage[latin9]{inputenc}
\usepackage{babel}
\usepackage{color}
\usepackage{amsmath}
\usepackage{graphicx}
\usepackage{amssymb}
\usepackage{esint}
\usepackage[unicode=true, pdfusetitle,
 bookmarks=true,bookmarksnumbered=false,bookmarksopen=false,
 breaklinks=false,pdfborder={0 0 1},backref=false,colorlinks=false]
 {hyperref}
\usepackage{amsmath}
\usepackage{amssymb}
\def\b{\begin{equation}} \def\e{\end{equation}}
\def\bd{\begin{displaystyle}} \def\ed{\end{displaystyle}}
\def\ba{\begin{array}} \def\ea{\end{array}}

\def\bee{\begin{enumerate}}
\def\eee{\end{enumerate}}

\def\1{\mbox{I\hspace{-.15em}1}}

\def\R{{\rm I\hspace{-.15em}R}}

\def\C{\hspace{3pt}{\rm l\hspace{-.47em}C}}

\def\b{\begin{equation}}
\def\e{\end{equation}}
\def\bee{\begin{enumerate}}
\def\eee{\end{enumerate}}

\makeatletter
\usepackage{latexsym}\usepackage{bm}

\makeatother

\begin{document}
\title{Super-gauge Field in de Sitter Universe}

\author{S. Parsamehr}
\email{sajad.parsamehr@srbiau.ac.ir} \affiliation{Department of
Physics, Science and Research Branch,\\ Islamic Azad University,
Tehran, Iran}
\author{M. Enayati}
\email{mohammad.enayati.61@gmail.com} \affiliation{Department of
Physics, Razi University, Kermanshah, Iran}
\date{\today}
\author{M.V. Takook}
\email{takook@razi.ac.ir} \affiliation{Department of Physics,
Science and Research Branch,\\ Islamic Azad University, Tehran,
Iran}

\begin{abstract}

\noindent \hspace{0.35cm} 

The Gupta-Bleuler triplet for vector-spinor gauge field is presented in de Sitter ambient space formalism. The invariant space of field equation solutions is obtained
with respect to an indecomposable representation of the de Sitter
group. By using the general
solution of the massless spin-$\frac{3}{2}$ field equation, the
vector-spinor quantum field operator and its corresponding Fock space is constructed. The quantum field operator can be written in terms of the vector-spinor
polarization states and a quantum conformally coupled massless
scalar field, which is constructed on Bunch-Davies vacuum state. The two-point function is also presented, which is de Sitter covariant and analytic.

\end{abstract}

\maketitle
\section{Introduction}

According to the highly redshift observation of the Supernova Ia
\cite{Perl, Riess}, galaxy clusters \cite{Henry 1, Henry 2}, and
cosmic microwave background radiation \cite{Nature}, the current
universe is expanding in an accelerating way. Then our current
universe may be described by the de Sitter space-time. Moreover, the
recently observational data by BICEP2 \cite{BICEP2} may confirm that
the early universe in a good approximation is also the de Sitter
universe. Therefore, the construction of the quantum field theory in
de Sitter space is very important for better understanding of the
evolution of the early and current universe. The rigorous
mathematical construction of quantum field theory in de Sitter
space-time, based on the unitary irreducible representations of the
de Sitter group and the analyticity of the complexified de Sitter
space-time, was previously presented in \cite{Takook 77}. The
unitary irreducible representations of the de Sitter group are
extracted completely by Takahashi \cite{TAKA} and the analyticity of
the complexified de Sitter space-time is investigated by Bros et al
\cite{Bros 1, Bros 2, Bros 3, Takook th}.

In this paper, the massless spin-$\frac{3}{2}$ field or
vector-spinor gauge field is considered. The massless means that
they propagate on the de Sitter light-cone. First by using the group
de Sitter algebra, the gauge invariant field equation is presented
\cite{Takook 77,Fatahi,azizi}. For similarity to other gauge
theories, such as Yang-Mills gauge theory, the gauge-covariant
derivative and the gauge invariant Lagrangian density can be
envisaged along the lines proposed in \cite{Takook 77}. The
variation of this Lagrangian density would give us a equation of
motion which is obtained by the group de Sitter algebra.

In the gauge-covariant derivative, the gauge potential is a
vector-spinor field. Consequently the corresponding gauge group must
have spinorial generator to justify a set of well-defined
gauge-covariant derivative. Therefore, a set of anti-commutative
generators satisfy a superalgebra.  The all possible closed de
Sitter superalgebra for even $N$ ($N$ is the number of fermionic
generators) had been obtained \cite{luno,Pilch}. In the de Sitter
ambient space notation, a closed $N=1$ de Sitter supersymmetry
algebra can be defined \cite{Pahlevan}. So, here we just consider
one spinorial generator and, in accordance with it, one
vector-spinor field. The quanta of this field is named gravitino
which is supposed to be the fermionic partner of graviton (spin-$2$
quanta of gravitational field!).

In the gauge quantum field theory, it has been shown that if we want
to conserve casuality and covariance, an indefinite metric must be
used \cite{Strocchi}. In other words, there are states with negative
or null norm that establish a general Fock quantization with field
operators that are not essentially self-adjoint \cite{Mintchev}, so,
one has to adopt the well-known Gupta-Bleuler quantization. The
Gupta-Bleuler formalism is an alternative way that is used by Gupta
and Bleuler to quantize the electromagnetic field \cite{Gupta,
Bleuler}. But it seems to be universal, and it has been extensively
applied to the quantization of gauge invariant theories. Binegar et
al. \cite{Fronsdal} have shown a complete Gupta-Bleuler quantization
procedure of QED that is manifestly conformally invariant. In a
curved static space-time, the Gupta-Bleuler quantization of the
electromagnetic gauge fields is explained by Furlani \cite{Furlani}
as well as for globally hyperbolic space-times in \cite{Finster}.
The Gupta-Bleuler structure has been applied to the massless
minimally coupled scalar field \cite{Bievre, Gazeau}, massless
vector field \cite{gagarota} and massless spin-2 field \cite{taro12}
in de Sitter space. Here, we study this structure for a massless
vector-spinor or spin-$\frac{3}{2}$ fields.

In section II, first, the notation and terminology of de Sitter
ambient space formalism are recalled. Using de Sitter group algebra,
the gauge invariant field equation is presented. The gauge-covariant
derivative is defined. Also we look closely at an action in which
its associated equation of motion, is exactly consistent with the
group algebra result \cite{Takook 77}. Section III is devoted to the
construction of Gupta-Bleuler triplet for vector-spinor field and
its corresponding indecomposable representation. The solution of the
gauge fixing field equation is obtained in section IV. The field
solution can be written in terms of a polarization vector-spinor
state and a conformally coupled massless scalar field. The pure
gauge state, spinor state, physical state, divergence part and
general solution are presented. In section V, the vector-spinor
quantum field operator and their covariant two-point function are
defined. Finally, a brief conclusion and an outlook are given in
section VI.


\setcounter{equation}{0}

\section{field equations}

\subsection{de Sitter ambient space formalism}

The de Sitter space-time is the vacuum solution of Einstein's
equation with a positive cosmological constant. It can be considered
as a hyperboloid embedded in five-dimensional Minkowski space:
\begin{equation*}
M_H=\{x \in \R^5| \; \; x \cdot x=\eta_{\alpha\beta} x^\alpha
x^\beta =-H^{-2}\},\;\;\alpha,\beta=0,1,2,3,4,
\end{equation*}
where $\eta_{\alpha\beta}=$diag$(1,-1,-1,-1,-1)$. The de Sitter
metric is
\begin{equation*}
ds^2=\eta_{\alpha\beta}dx^{\alpha}dx^{\beta}|_{x^2=-H^{-2}}=
g_{\mu\nu}^{dS}dX^{\mu}dX^{\nu},\;\; \mu=0,1,2,3.
\end{equation*}
$X^\mu$ is de Sitter intrinsic coordinates and $x^\alpha$ is the
five-dimensional de Sitter ambient space formalism. For simplicity,
the Hubble parameter is taken to be equal to unit $H=1$. The
isometry group of the de Sitter space-time is the ten-parameter
group $SO_0(1,4)$. The de Sitter group has two Casimir operators:
\begin{equation*}
Q^{(1)}=-\frac{1}{2}L_{\alpha\beta}L^{\alpha\beta},
\;\;\;  \mbox{and} \;\;\;
Q^{(2)}=-W_\alpha W^\alpha,
\end{equation*}
where
$ W_\alpha =\frac{1}{8}\epsilon_{\alpha\beta\gamma\delta\eta}
L^{\beta\gamma}L^{\delta\eta}. $
$\epsilon_{\alpha\beta\gamma\delta\eta}$ is the antisymmetrical
Levi-Civita tensor and
$L_{\alpha\beta}=M_{\alpha\beta}+S_{\alpha\beta}$ are ten
infinitesimal generators of de Sitter group. The orbital part
$M_{\alpha\beta}$ is defined by
\begin{equation*}
M_{\alpha \beta}=-i(x_\alpha \partial_\beta-x_\beta
\partial_\alpha)=-i(x_\alpha\partial^\top_\beta-x_\beta
\partial^\top_\alpha),
\end{equation*}
where $\partial^\top_\beta=\theta_\beta^{\;\;\alpha}\partial_\alpha$
is the transverse derivative ($x.\partial^\top=0$) and $\theta
_{\alpha\beta}=\eta_{\alpha\beta}+x_\alpha x_\beta$ known as the
projection tensor. The
spinorial part $S_{\alpha\beta}$ with half-integer spin
$s=l+\frac{1}{2}$, is read as
\begin{equation*}
S_{\alpha\beta}^{(s)}=S_{\alpha\beta}^{(l)}+S_{\alpha\beta}^{(\frac{1}{2})},
\end{equation*}
in which the first term acts on a tensor index as
\begin{equation*}
S_{\alpha\beta}^{(l)}\Psi_{\gamma_1...\gamma_l}=-i\sum\limits_{i=1}^l\left(\eta_{\alpha{\gamma_i}}
\Psi_{\gamma_1...(\gamma_i\rightarrow\beta)...\gamma_l}-\eta_{\beta{\gamma_i}}
\Psi_{\gamma_1...(\gamma_i\rightarrow\alpha)...\gamma_l}\right).
\end{equation*}
The second term is
\begin{equation*}
S_{\alpha\beta}^{(\frac{1}{2})}=-\frac{i}{4}[\gamma_{\alpha},\gamma_{\beta}],
\end{equation*}
where the $\gamma$-matrices satisfy the basic Clifford
algebra
\begin{equation*}
\{\gamma^{\alpha},\gamma^{\beta}\}=2\eta^{\alpha\beta}\mathbb{I}_{4\times4}.
\end{equation*}
The best $\gamma$-matrices representation for our discussion is
\cite{Takook th,sep}:
\begin{equation*}
\gamma^0=\left( \begin{array}{clcr} \mathbb{I}_{2\times2} & \;\; 0 \\ 0 &-\mathbb{I}_{2\times2} \\
\end{array}\right)
,\gamma^4=\left( \begin{array}{clcr} 0 & \mathbb{I}_{2\times2} \\ -\mathbb{I}_{2\times2} &0 \\
\end{array}\right),
\end{equation*}
\begin{equation*}
\gamma^1=\left( \begin{array}{clcr} 0 & i\sigma^1 \\ i\sigma^1 &0 \\
\end{array} \right)
,\gamma^2=\left( \begin{array}{clcr} 0 & -i\sigma^2 \\ -i\sigma^2 &0 \\
\end{array} \right)
,\gamma^3=\left( \begin{array}{clcr} 0 & i\sigma^3 \\ i\sigma^3 &0 \\
\end{array} \right),
\end{equation*}
where $\mathbb{I}_{2\times2}$ and $\sigma^i$'s are unit $2\times 2$ matrix and
the Pauli matrices, respectively.

\subsection{Gauge invariant equation}

The field equation can be written by using the
second-order Casimir operator of de Sitter group \cite{Takook 77}:
\begin{equation}\label{II.3}
\left(Q_{j,p}^{(1)}-\left<Q_{j,p}^{(1)}\right>\right)\Psi(x)=0,
\end{equation}
where the eigenvalues of Casimir operator, which classify the
unitary irreducible representations of de Sitter group, is
\begin{equation*}
\left<Q_{j,p}^{(1)}\right>=\left(-j(j+1)-(p+1)(p-2)\right).
\end{equation*}
$j$ and $p$ are parameters in which they take values corresponding
to different types of representations, namely: the principal
$(U^{(j,p)})$, the complementary $(V^{(j,p)})$, and the discrete
series $(\Pi^\pm_{j,p})$. A la Wigner, the quantum field operator
transforms by the unitary irreducible representations of de Sitter
group.

In the following, we will concentrate on the spin-$\frac{3}{2}$
massless vector-spinor field corresponding to the
 values $j=p=\frac{3}{2}$ in the discrete series
with $\left<Q_{\frac{3}{2},\frac{3}{2}}^{(1)}\right>=-\frac{5}{2}$.
Therefore, \eqref{II.3} becomes
\begin{equation}\label{II.4}
\left(Q_{\frac{3}{2}}^{(1)}+\frac{5}{2}\right) \Psi_\alpha=0,
\end{equation}
where $Q_{j,p}^{(1)}\equiv Q_{j}^{(1)}$ and \cite{Takook th}
\begin{equation*}
Q_{\frac{3}{2}}^{(1)}\Psi_\alpha=Q_0^{(1)}\Psi_\alpha+\slashed{x}\slashed{\partial}^\top\Psi_\alpha
+2x_\alpha\partial^\top\cdot\Psi-\frac{11}{2}\Psi_\alpha+\gamma_\alpha
\slashed{\Psi}.
\end{equation*}
The "scalar" Casimir operator $Q_0^{(1)}$ is:
\begin{equation*}
Q_0^{(1)}=-\frac{1}{2}M^{\alpha\beta}M_{\alpha\beta}=-\partial_\alpha^\top\partial^{\alpha\top}.
\end{equation*}
The vector-spinor solution of the field equation \eqref{II.4} with
the condition $\partial^\top.\Psi=0$ is  singular \cite{TAB}. This
condition is necessary for the transformation of the field operator
by an unitary irreducible representation of de Sitter group. One can
solve the problem of singularity by release of the
divergencelessness condition, i.e. $\partial^\top.\Psi\neq0$. Then
the quantum field operator transforms by an indecomposable
representation of de Sitter group and the field equation must be
gauge invariant \cite{Takook 77}. The massless vector-spinor gauge
invariant field equation is \cite{Takook 77,Fatahi,azizi}:
\begin{equation}\label{II.5}
\left(Q_{\frac{3}{2}}^{(1)}+\frac{5}{2}\right) \Psi_\alpha+
\nabla_{\alpha}^\top\partial^\top\cdot\Psi=0.
\end{equation}
$\nabla_{\alpha}^\top$ is a transverse-covariant derivative which
maps a tensor-spinor field of rank $l$ to a tensor-spinor field of
rank $l+1$ on the de Sitter ambient space formalism \cite{Takook
77}:
\begin{equation*}
\nabla^\top_\beta \Psi_{\alpha_1....\alpha_l}\equiv
(\partial^\top_\beta+\gamma^\top_\beta\slashed{x})
\Psi_{\alpha_1....\alpha_l}-\sum_{n=1}^{l}x_{\alpha_n}\Psi_{\alpha_1..\alpha_{n-1}\beta\alpha_{n+1}..\alpha_l},
\end{equation*}
where $\slashed{x}=\gamma_\alpha x^\alpha$ and
$\gamma^\top_\alpha=\theta_{\alpha}^{\beta}\gamma_\beta$. It is
clear that if someone eliminates $\gamma^\top_\beta\slashed{x}$ from
the above definition, the transverse-covariant derivative will be
transverse again. But, the de Sitter algebra and the definition of
the Casimir operator persuade us to add this term \cite{Takook 77}.
Then one can prove the following identities:
\begin{equation}\label{II.6}
Q^{(1)}_{\frac{3}{2}}\nabla_{\alpha}^\top =\nabla_{\alpha}^\top
Q^{(1)}_{\frac{1}{2}}\:\:\:,\:\:\:\partial^\top\cdot \nabla^\top
\psi=-\left(Q^{(1)}_{\frac{1}{2}}+ \frac{5}{2}\right)\psi.
\end{equation}
By using these identities, one can show that the field equation
\eqref{II.5} is invariant under the
following gauge transformation
\begin{equation*}
\Psi_\alpha \longrightarrow
\Psi^g_\alpha=\Psi_\alpha+\nabla_{\alpha}^\top \psi.
\end{equation*}
$\psi$ is an arbitrary spinor field and
\begin{equation*}
Q_{\frac{1}{2}}^{(1)}\psi=\left( Q_0+\slashed{
x}\slashed{\partial}^\top-\frac{5}{2}\right)\psi.
\end{equation*}

\subsection{Gauge-covariant derivative}

The local or gauge symmetries are fundamental in the nature to
explain the electromagnetic, weak and strong nuclear interactions by
the gauge vector fields. To construct a gauge-invariant Lagrangian,
the gauge-covariant derivative is defined such that any gauge fields
are associated with the generators of the local symmetry group.

Here,  the gauge field is vector-spinor field ($\Psi_\alpha$) which
satisfies an anti-commuting algebra, then the associated generator
must be an spinor and satisfy the anti-commutation relations.  In
this case, the super-gauge covariant derivative is defined by
\cite{Takook 77}
\begin{equation*}
D^\Psi_\beta \equiv\nabla_{\beta}^\top +i \left(\Psi_{\beta}\right)^\dag
\gamma^0 {\cal Q}=\nabla_{\beta}^\top
+i\left(-\bar{\Psi}_{\beta}\gamma^4\right)^{i} {\cal Q}_i,
\end{equation*}
where $\cal Q$ is a fermionic generator which transforms as spinor
under the de Sitter group \cite{Pahlevan}.
$\bar{\Psi}_{\beta}=\Psi_{\beta}^\dag \gamma^0\gamma^4$and
$i=1,...,4$ is the spinorial index. By this fermionic
generator, one can not define an closed algebra. It was proved that
this spinor generator, $\cal Q$, with the de Sitter group algebra
satisfy the following $N=1$ supersymmetric de Sitter algebra in
ambient space formalism \cite{Pahlevan}
\begin{equation*}
\{{\cal Q}_i,{\cal
Q}_j\}=\left(S^{(\frac{1}{2})}_{\alpha\beta}\gamma^4
\gamma^2\right)_{ij}L^{\alpha\beta},
\end{equation*}
\begin{equation*}
[{\cal
Q}_i,L_{\alpha\beta}]=\left(S^{(\frac{1}{2})}_{\alpha\beta}{\cal Q}
\right)_i, \;\; [\widehat {\cal Q}_i,L_{\alpha\beta}]=-\left(\widehat
{\cal Q} S^{(\frac{1}{2})}_{\alpha\beta}\right)_i,
\end{equation*}
\begin{equation} \label{dsal}
[L_{\alpha\beta}, L_{\gamma\delta}] =
-i(\eta_{\alpha\gamma}L_{\beta\delta}+\eta_{\beta\delta}
L_{\alpha\gamma}-\eta_{\alpha\delta}L_{\beta\gamma}-\eta_{\beta\gamma}
L_{\alpha\delta}),
\end{equation}
where $\widehat{{\cal Q}}_i\equiv\left({\cal Q}^t \gamma^4 C\right)_i$.
${\cal Q}^t$ is the transpose of $\cal Q$ and $C$ is the charge
conjugation \cite{TRS}. One can prove that $\widehat{{\cal
Q}}\gamma^4\cal Q $ is a scalar field under de Sitter group
transformations \cite{TRS}.

Therefore naturally the vector-spinor gauge field $\Psi_{\beta}$
(associated to the fermionic generator ${\cal Q}$) must be coupled
with a tensor gauge field ${\cal K}_{\beta}^{\;\; \gamma\delta}$
(associated to the generator $L_{\gamma\delta}$) \cite{Takook 77}.
${\cal K}_{\beta}^{\;\; \gamma\delta}$ is a massless spin-$2$
rank-$3$ mixed-symmetric tensor field \cite{Takook 77}. In this
case, the general gauge-covariant derivative, with ${\cal
H}_\alpha^{\;\;A}\equiv  \left( {\cal K}_{\beta}^{\;\;
\gamma\delta}, \Psi_{\beta}^i\right)$ as gauge fields, can be
defined as:
\begin{equation*}
D^{\cal H}_\beta =\nabla^\top_\beta +i {\cal H}_{\beta}^{\;\; A}
T_A,
\end{equation*}
where $ T_A \equiv  \left( L_{\alpha\beta}, {\cal Q}_i\right)$ are
the generators of $N=1$ super-de Sitter algebra (\ref{dsal}). For
simplicity, they can be written in the following compact form:
\begin{equation*}
[T_A,T_B\}={\cal C}_{BA}^{\;\;\;\;\;C}T_C.
\end{equation*}
The symbol of $[ $ $\}$ is an anti-commutation if and only if the
two $T$'s are fermionic; otherwise, it is a commutation symbol. A
general form of local infinitesimal gauge transformation acting on
the gauge field can be written as
\begin{equation*}
\delta_\epsilon {\cal H}_\beta^{\;\;A}=D^{\cal H}_\beta \epsilon^A
=\nabla^\top_\beta \epsilon^A + {\cal C}_{BC}^{\;\;\;\;\;A}{\cal
H}_\beta^{\;\;C} \epsilon^B.
\end{equation*}
According to the general framework, one can obtain
\begin{equation*}
[D^{\cal H}_\alpha,D^{\cal H}_\beta\
\}=R_{\alpha\beta}^{\;\;\;\;A}T_A,
\end{equation*}
where $R$ is the "curvature" and is defined as
\begin{equation*}
R_{\alpha\beta}^{\;\;\;\;A}=\nabla^\top_\alpha {\cal
H}_\beta^{\;\;A}-\nabla^\top_\beta {\cal H}_\alpha^{\;\;A}+{\cal
H}_\beta^{\;\;B}{\cal H}_\alpha^{\;\;C}{\cal
C}_{BC}^{\;\;\;\;\;A},\;\;\; x^\alpha R_{\alpha\beta}^{\;\;\;\;A}=0=
x^\beta R_{\alpha\beta}^{\;\;\;\;A}.
\end{equation*}

Here we only consider the vector-spinor field part, then the
curvature for this part is:
\begin{equation*}
R_{\alpha\beta}^{\;\;\;\;\; i}=\nabla^\top_\alpha
\Psi_\beta^{\;\;i}-\nabla^\top_\beta \Psi_\alpha^{\;\;i} +{\cal
H}_\beta^{\;\;B}{\cal H}_\alpha^{\;\;C}{\cal C}_{BC}^{\;\;\;\;\;i},
\end{equation*}
where the transverse-covariant derivative acts on $\Psi_\beta$ in
the following form
\begin{equation*}
\nabla^\top_\alpha \Psi_\beta= \partial^\top_\alpha \Psi_\beta+
\gamma^\top_\alpha \slashed{x}\Psi_\beta -x_\beta \Psi_\alpha.
\end{equation*}

\subsection{Gauge invariant Lagrangian density}

The super-gauge invariant action or the supergravity Lagrangian in
the de Sitter ambient space formalism is \cite{MM,Takook 77}:
\begin{equation*}
S_g=\int
d\mu(x) R_{\alpha\beta}^{\;\;\;\;A}g_{AB}
R^{\alpha\beta B},
\end{equation*}
where $g_{AB}$ is numerical constant matrix and $d\mu(x)$ is the de Sitter invariant volume element \cite{Bros 2}. For vector-spinor field part, the action is given by
\begin{equation*}
S_g[\Psi,{\cal K}]=\int d\mu(x)
\left(\tilde{R}^i\right)_{\alpha\beta
}\left(R^i\right)^{\alpha\beta },
\end{equation*}
where
\begin{equation*}
\tilde{R}^i_{\alpha\beta}=\tilde{
\nabla}^\top_\alpha \tilde{\Psi}_\beta^{\;\;i}
-\tilde{\nabla}^\top_\beta \tilde{\Psi}_\alpha^{\;\;i} +{\cal
H}_\beta^{\;\;{\cal B}}{\cal H}_\alpha^{\;\;{\cal C}}{\cal C}_{{\cal
B}{\cal C}}^{\;\;\;\;\;i}.
\end{equation*}
The conjugate spinor is defined as $\tilde\Psi_{\alpha}\equiv
\Psi^\dag_{\alpha}\gamma^0$ and its transverse covariant-derivative
must be defined as \cite{Takook 77}:
\begin{equation}\label{II.2}
\tilde{ \nabla}^\top_\beta \tilde{\Psi}_{\alpha}\equiv
\partial^\top_\beta
\tilde{\Psi}_{\alpha}-x_{\alpha}\tilde{\Psi}_{\beta}, \;\;\; \tilde \nabla_\alpha^\top \tilde\psi =\partial_\alpha^\top
\tilde\psi.
\end{equation}

In the approximation of the linear field equation, the action is
\begin{equation} \label{action}
S[\Psi, \tilde{\Psi}]\simeq \int
d\mu(x)\left[\left(\tilde{\nabla}^\top_\alpha\tilde{
\Psi}_\beta-\tilde{\nabla}^\top_\beta\tilde{
\Psi}_\alpha\right)\left( \nabla^{\top\alpha}
\Psi^\beta-\nabla^{\top\beta} \Psi^\alpha\right)\right].
\end{equation}
Using the Euler-Lagrange equation, the field equations for two dynamical variables $\Psi$
and $\tilde\Psi$, can be obtained as [Appendix A]:
\begin{equation} \label{vsfe0}
(x_\alpha-\partial^{\top}_{\alpha}) \left(\nabla^{\top\alpha}
\Psi^{\beta}-\nabla^{\top\beta} \Psi^{\alpha} \right)=0,
\end{equation}
\begin{equation}
\left(\partial^{\top\alpha}-\gamma^\alpha\slashed{x}\right)
\left(\tilde{\nabla}^\top_\alpha\tilde{
\Psi}_\beta-\tilde{\nabla}^\top_\beta\tilde{ \Psi}_\alpha\right)=0.
\end{equation}
The above equations of motion in
terms of Casimir operator can be rewritten as the following forms:
\begin{equation}\label{II.10}
\left(Q_{\frac{3}{2}}^{(1)}+\frac{5}{2}\right) \Psi_\alpha+
\nabla_{\alpha}^\top\partial^\top\cdot\Psi=0,
\end{equation}
\begin{equation}
\left(Q_{\frac{3}{2}}^{(1)}+\frac{5}{2}\right)\tilde
\Psi_\alpha+\partial_{\alpha}^\top\left(\slashed{x}\tilde{\slashed{\Psi}}
+\partial^\top\cdot\tilde\Psi\right)-2\left(\gamma_\alpha\tilde{\slashed{\Psi}}
+\slashed{x}\slashed{\partial}^\top\tilde\Psi_\alpha-\tilde\Psi_\alpha\right)=0.
\end{equation}
It is also shown that \eqref{II.10} is completely consistent with
equation \eqref{II.5} that is calculated on the basis of the group
theory approach. The vector-spinor Lagrangian density is then
invariant under the following gauge transformations [Appendix
B]:
\begin{equation*}
\Psi_\alpha \longrightarrow
\Psi^g_\alpha=\Psi_\alpha+\nabla_{\alpha}^\top  \psi,
\end{equation*}
\begin{equation*}
\tilde{\Psi}_\alpha \longrightarrow
\tilde{\Psi}^g_\alpha=\tilde{\Psi}_\alpha+\partial_{\alpha}^\top
\tilde{\psi}.
\end{equation*}

We would like to introduce a gauge fixing parameter $c$:
\begin{equation}\label{II.12}
\left(Q_{\frac{3}{2}}^{(1)}+\frac{5}{2}\right) \Psi_\alpha+
c\nabla_{\alpha}^\top \partial^\top\cdot\Psi=0.
\end{equation}
The equation \eqref{II.5}, which is a gauge invariant equation, is a
special case of the above equation. When $c\neq1$, we have a field
equation which is not gauge invariant. The choice of the gauge
fixing parameter $c$ determines the space of gauge solutions, which
will be considered in the next section.

\setcounter{equation}{0}
\section{Gupta-Bleuler triplet}

The appearance of the Gupta-Bleuler triplet is crucial for the
covariant quantization of the gauge fields
\cite{Strocchi,Fronsdal,gagarota}. The ambient space formalism
allows us to exhibit this triplet for the vector-spinor gauge field
in exactly the same manner as it occurs for the Minkowskian
counterpart. We start with the gauge fixing field equation
(\ref{II.12}). The de Sitter invariant bilinear form (or inner
product) on the space of solutions is defined for two modes of the
field equation (\ref{II.12}). Let us now define the structure of the
space of solutions as the Gupta-Bleuler triplet $V_g \subset V
\subset V_c$.

The indefinite inner product space $V_c$ includes all the solutions
of the field equation \eqref{II.12}. In other words, the elements of
this space are physical and unphysical states with all possible
norms such as negative, null, and positive. The subspace $V$ is
defined as a space of solution with the divergencelessness
condition, $\partial^\top \cdot \Psi=0$. This subspace $V$ is a
semi-definite inner product space and an invariant subspace of $V_c$
(but not invariantly complemented). According to \eqref{II.12}, it
is a manifestly $c$-independent subspace of solutions. Finally, the
gauge subspace $V_g \subset V$ is defined as
$\Psi_\alpha^g=\nabla_\alpha^\top\psi^p$, where $p$ stands for pure
gauge state. It establishes a subspace with the null norm, which is
an invariant subspace of $V$ (but not invariantly complemented). The
elements of $V_g$ are orthogonal to all states in $V$ including
themselves. The coset space $V/V_g$ is the space of the physical
state. In the following, we present these three spaces.

\subsection{The pure gauge state}

Putting the gauge solution $\Psi_\alpha^g =\nabla_{\alpha}^\top\psi^p$ into \eqref{II.12} and
by using \eqref{II.6}, we obtain:
\begin{equation}
(1-c)\nabla_{\alpha}^\top\left(Q_{\frac{1}{2}}^{(1)}+\frac{5}{2}\right)\psi^p
=(1-c)\nabla_{\alpha}^\top(Q_0+\slashed{x}\slashed{\partial}^\top)\psi^p=0,
\end{equation}
where $\psi^p$ is a spinor field.
We will make the following assumptions:

\begin{itemize}

\item  If $c=1$, the spinor field $\psi^p$ is arbitrary and unlimited. The field equation is gauge invariant.
The gauge vector-spinor space is constructed by a spinor field $\psi^p$.

\item  If  $c\neq1$, then $\psi^p$ obeys the following field equation
       \begin{equation*}
       \nabla_{\alpha}^\top\left(Q_{\frac{1}{2}}^{(1)}+\frac{5}{2}\right)\psi^p=0,
       \end{equation*}
       or, for simplicity, we can chose
       \begin{equation}\label{III.2}
       \left(Q_{\frac{1}{2}}^{(1)}+\frac{5}{2}\right)\psi^p=(Q_0+\slashed{x}\slashed{\partial}^\top)\psi^p=0.
       \end{equation}
       In this case, the gauge is fixing and the field equation is not a gauge invariant.
\end{itemize}
This field can be associated to the de Sitter group representation
$\Pi_{\frac{1}{2},{-\frac{1}{2}}}$.

\subsection{The divergence spinor state}

The divergence vector-spinor state is defined as $\partial^\top\cdot\Psi^d \neq 0$. If one takes the
divergence of the field equation \eqref{II.12},
\begin{equation*}
\partial^{\top\alpha}\left
[\left(Q_\frac{3}{2}^{(1)}+\frac{5}{2}\right)\Psi^d_\alpha+c\nabla_{\alpha}^\top
\partial^\top\cdot\Psi^d\right ]=0,
\end{equation*}
then one obtain [Appendix C]:
\begin{equation*}
(1-c)\left(Q_{\frac{1}{2}}^{(1)}+\frac{5}{2}\right)\partial^\top\cdot\Psi^d=(1-c)(Q_0+\slashed{x}\slashed{\partial}^\top)\partial^\top\cdot\Psi^d=0.
\end{equation*}
At this point, one must consider the two cases $c=1$ and $c\neq1$:
\begin{itemize}
\item
        If $c=1$, $\partial^\top\cdot\Psi^d\equiv\psi^s$, where $s$ stands for spinor state, $\psi^s$ is an arbitrary spinor field and we have a gauge invariant.
\item
        If  $c\neq1$, then $\psi^s$ satisfies the following
        field equation:
        \begin{equation}\label{III.3}
        \left(Q_{\frac{1}{2}}^{(1)}+\frac{5}{2}\right)\psi^s=0,
        \end{equation}
        and the gauge is fixed.
\end{itemize}
The quotient space$V_c/V$is the space of spinor states $\psi^s$.
This field, similar to the pure gauge state, can be associated with
the representation $\Pi_{\frac{1}{2},{-\frac{1}{2}}}$.

\subsection{The physical state}

The quotient space $V/V_g$ is the space of the physical states.
These states are the solutions of the field equation \eqref{II.4}
with the conditions: $\partial^\top\cdot\Psi^{phy}=0$, $\gamma\cdot
\Psi^{phy}=0$ and $\Psi_\alpha^{phy}\neq\nabla_{\alpha}^\top\psi$.
These fields are transformed by the discrete series representation
$\Pi^\pm_{\frac{3}{2},{\frac{3}{2}}}$.

\vspace{5mm}
In this way, we obtain approximately what is known as
the indecomposable group representation structure for the massless
vector-spinor field
\begin{equation*}
\underbrace{\Pi_{\frac{1}{2},-\frac{1}{2}}}_{spinor \;
representation}\;\;\longrightarrow\underbrace{\Pi_{\frac{3}{2},\frac{3}{2}}^{+}
\oplus\Pi_{\frac{3}{2},\frac{3}{2}}^{-}}_{physical \;
representation}\longrightarrow\;\;\underbrace{\Pi_{\frac{1}{2},-\frac{1}{2}}}_{pure
\; gauge \; representation}
\end{equation*}
where the arrows indicate the state leak under the group action. The
spin-$\frac{3}{2}$ unitary irreducible representations of the de
Sitter group with the helicity $\pm\frac{3}{2}$ get involved in the
central part $\Pi^\pm_{\frac{3}{2},{\frac{3}{2}}}$ and one can see
that they contract to the Poincar\'{e} massless spin-$\frac{3}{2}$
representations when the curvature tends to zero \cite{Takook 77}. As
we can see, the spinor and pure gauge states are associated with the
representation $\Pi_{\frac{1}{2},-\frac{1}{2}}$.

\setcounter{equation}{0}
\section{Field solution}

For simplicity, the vector-spinor field solution is divided into three parts:
\begin{equation*}
\Psi_\alpha(x)=\Psi_\alpha^g+\Psi_\alpha^d+\Psi_\alpha^{phy},
\end{equation*}
where $\Psi_\alpha^g$ is the pure gauge solution, $\Psi_\alpha^d$ is
the divergence part solution and $\Psi_\alpha^{phy}$ is the physical
solution. We have defined $\Psi_\alpha^g=\nabla_\alpha^\top\psi^p$,
so, if one takes the divergence of it, one obtain:
\begin{equation*}
\partial^\top\cdot\Psi^g=-\left(Q_0+\slashed{x}\slashed{\partial}^\top\right)\psi^p.
\end{equation*}
The gauge solution
satisfies the divergencelessness condition $\partial^\top\cdot\Psi^g=0$. Then the spinor field equation is \eqref{III.2}. Also, one can not impose
the condition $\slashed{\Psi}^g=0$ for spinor and gauge state due to
the homogeneous degree of spinor field states (see equation
(\ref{unphysical})). From the unitary irreducible representations of
the de Sitter group, we know that the physical solution must satisfy
the conditions $\partial^\top\cdot\Psi^{phy}=0$ and
$\slashed{\Psi}^{phy}=0$. Therefore, the only divergence part is
$\partial^\top\cdot\Psi^d \neq 0$. For a classification, see
Table.I.

\begin{table}[ht]

\begin{tabular}{c c c c}
\hline
State & Field equation & Condition I & Condition II\\[0.5ex]
\hline\hline
Physical State: & $\left(Q_{\frac{3}{2}}^{(1)}+\frac{5}{2}\right)\Psi_\alpha^{phy}=0,$ & $\partial^\top \cdot \Psi^{phy}=0,$ & $\slashed{\Psi}^{phy}=0,$
\\[2ex]
Pure Gauge State: & $\left(Q_{\frac{3}{2}}^{(1)}+\frac{5}{2}\right)\Psi_\alpha^g=0,$ & $\partial^\top \cdot \Psi^g=0,$ & $\slashed{\Psi}^g \neq 0,$
\\[2ex]
Divergence State: & $\left(Q_{\frac{3}{2}}^{(1)}+\frac{5}{2}\right)\Psi_\alpha^d+c\nabla_\alpha^\top \partial^\top \cdot \Psi^d=0,$ & $\partial^\top \cdot \Psi^d \neq 0,$ & $\slashed{\Psi}^d=0.$ \\
[2ex] \hline
\end{tabular}
\caption{The Gupta-Bleuler Triplet States}
\centering
\label{table:nonlin}
\end{table}

\subsection{The pure gauge and divergence spinor solution}

The pure gauge field $\psi^{p}$ and the spinor state $\psi^{s}$
satisfy the similar field equations, \eqref{III.2} and
\eqref{III.3}:
\begin{equation}\label{IV.1}
\left(Q_{\frac{1}{2}}^{(1)}+\frac{5}{2}\right)\psi^{p,s}=0, \;\;\;
or \;\;\;
\left(Q_0^{(1)}+\slashed{x}\slashed{\partial}^\top\right)\psi^{p,s}=0.
\end{equation}
By using the identity
\begin{equation}\label{q0iden}
Q_0^{(1)}=\left(3-\slashed{x}\slashed{\partial}^\top\right)\slashed{x}\slashed{\partial}^\top
=\slashed{x}\slashed{\partial}^\top\left(3-\slashed{x}\slashed{\partial}^\top\right),
\end{equation}
there exist two possibilities for the first-order field equation.
The first one is:
\begin{equation}\label{minspinor}
\slashed{x}\slashed{\partial}^\top \psi^{p,s}=0, \;\;\;
Q_0^{(1)}\psi^{p,s}=0,
\end{equation}
with the degree of homogeneity $-3$ and $0$ \cite{Takook 77}. The other field
equation reads as:
\begin{equation}\label{unphysical}
\left(\slashed{x}\slashed{\partial}^\top-4\right) \psi^{p,s}=0,
\;\;\; \left(Q_0^{(1)}+4\right)\psi^{p,s}=0,
\end{equation}
with the degree of homogeneity $-4$ and $1$. For the second case,
due to the positive homogeneous degree $1$, one can not construct
a covariant quantum field operator \cite{Takook 77}.

The solution of the field equation \eqref{minspinor} can be written
in the following form:
\begin{equation}\label{pgsm}
\psi^{p,s}=\left(3-\slashed{x}\slashed{\partial}^\top\right){\cal U}
\phi_m,
\end{equation}
where $\phi_m$ is a massless minimally scalar field
($Q_0^{(1)}\phi_m=0$). ${\cal U}$ is an arbitrary constant spinor
which can be fixed by imposing the condition that it becomes the spinor field in the null curvature limit
\cite{Takook th, sep}. The solution of the massless minimally
coupled scalar field can be written in terms of the de Sitter
plane-wave: $(x\cdot \xi)^\sigma$ \cite{Bros 2, Takook 77}, where
$\xi$ is an $5$-vector in positive cone ${\cal C}^+$:
\begin{equation} \label{poscon}
\xi \in {\cal C}^+=\{ \xi \in R^5;\;\; \eta_{\alpha \beta}\xi^\alpha
\xi^\beta=(\xi^0)^2-\vec \xi\cdot\vec \xi-(\xi^4)^2=0,\; \xi^0>0 \}.
\end{equation}
For the massless minimally coupled scalar field, the degree of
homogeneity is $\sigma=0,-3$. The constant solution poses the famous
zero mode problem for this field. By using the following relation
between the minimally coupled and the conformally coupled scalar
fields $((Q_0-2)\phi_c=0)$ in the de Sitter ambient space formalism
\cite{Takook 77}
\begin{equation}\label{IV.11}
\phi_m=\left[Z\cdot\partial^\top+2Z\cdot x\right]\phi_c,
\end{equation}
this problem can be surmounted \cite{taro15}. $Z_\alpha$ is a constant
 five-vector. The solution of the massless conformally
coupled scalar field can be written in terms of the de Sitter
plane-wave: $(x\cdot \xi)^\sigma$, $\sigma=-1,-2$ \cite{Bros 2,
Takook 77}. Then the spinor field (\ref{pgsm}) can be written in
terms of the massless conformally coupled scalar field as:
\begin{equation}\label{pgsc}
\psi^{p,s}=\left(3-\slashed{x}\slashed{\partial}^\top\right){\cal U}
\left[Z\cdot\partial^\top+2Z\cdot x\right]\phi_c.
\end{equation}
There appear an arbitrary constant spinor ${\cal U}$ and an arbitrary constant
 five-vector $Z^\alpha$, which will be fixed in the null curvature limit.

\subsection{The physical state solution}

The physical part, which is defined by the conditions $\partial^\top
\cdot \Psi_{\alpha}^{phy}=0$ and $\slashed{\Psi}^{phy}=0$, satisfies
the following field equation:
\begin{equation*}
\left(Q^{(1)}_{\frac{3}{2}}+\frac{5}{2}\right)\Psi_{\alpha}^{phy}
=\left(\slashed{x}\slashed{\partial}^\top-3\right)\left(-\slashed{x}\slashed{\partial}^\top+1\right)\Psi_{\alpha}^{phy}=0.
\end{equation*}
There are two possibilities for the relevant first order field
equation:
\begin{equation*}
\left(\slashed{x}\slashed{\partial}^\top-3\right)\Psi_{\alpha}^{phy}=0,\;\;\;Q^{(1)}_0\Psi_{\alpha}^{phy}=0,
\end{equation*}
and
\begin{equation*}
\left(\slashed{x}\slashed{\partial}^\top-1\right)\Psi_{\alpha}^{phy}=0,\;\;\;\left(Q^{(1)}_0-2\right)\Psi_{\alpha}^{phy}=0.
\end{equation*}
The latter is conformal invariant \cite{Fatahi}, and in the
following, only this solution will be considered. The physical
vector-spinor field solution can be written in terms of a
polarization vector-spinor $\textbf{D}^{phy}_\alpha$ and a spinor
field $\psi_1$ \cite{Fatahi}:
\begin{equation*}
\Psi_\alpha^{phy}=\textbf{D}^{phy}_\alpha(x,\partial^\top,Z)\psi_1,
\end{equation*}
where
\begin{equation*}
\textbf{D}^{phy}_\alpha(x,\partial^\top,Z)\equiv
Z_\alpha^\top+\left[\frac{1}{2}\nabla_\alpha^\top(1+3\slashed{x})
-\frac{5}{4}\gamma_\alpha^\top(1-\slashed{x})\right]\left[\slashed{Z}+(1+3\slashed{x})x.Z\right].
\end{equation*}

The spinor field $\psi_1$ satisfies
\begin{equation*}
\left(\slashed{x}\slashed{\partial}^\top-1\right)\psi_1=0,\;\;\;\left(Q^{(1)}_{\frac{1}{2}}-\frac{1}{2}\right)\psi_1=\left(Q^{(1)}_0-2\right)\psi_1=0.
\end{equation*}
Its solution can be written in terms of a massless conformally
coupled scalar field $\phi_c$ as \cite{Fatahi}:
\begin{equation}\label{IV.6}
\psi_1=\left(2-\slashed{x}\slashed{\partial}^\top\right){\cal
U}\phi_c.
\end{equation}
This spinor field and its related two-point functions can in fact be
extracted from a massive spinor field in the principal series
representation by setting $\nu=-i$ \cite{sep}.

\subsection{$\Psi_{\alpha}^{d}$ solution with $c=\frac{2}{3}$}

The divergence part is defined as $\partial^\top\cdot
\Psi_{\alpha}=\partial^\top \cdot \Psi_{\alpha}^{d}\neq 0$ and
$\slashed{\Psi}^{d}= 0 $. It satisfies the field equation:
\begin{equation} \label{dp}
\left(Q^{(1)}_{\frac{3}{2}}+\frac{5}{2}\right)\Psi_{\alpha}^d+c\nabla_\alpha^\top\partial^\top\cdot\Psi^d=0.
\end{equation}
This vector-spinor field $\Psi_{\alpha}^{d}$ can be expressed
by three spinor fields $\zeta_1,\zeta_2$ and $\zeta_3$ as follows
\cite{Fatahi}:
\begin{equation} \label{dps}
\Psi_\alpha^{d}=Z_{\alpha}^\top\zeta_1+\nabla_\alpha^\top\zeta_2+\gamma_\alpha^\top\zeta_3.
\end{equation}
By replacing (\ref{dps}) in the field equation (\ref{dp}), one
obtains:
\begin{equation}\label{upe1}
\left(Q_0+\slashed{x}\slashed{\partial}^\top-3\right)\zeta_1=0,
\end{equation}
\begin{equation}\label{upe2}
\slashed{x}x.Z\zeta_1-\slashed{x}\left(4-\slashed{x}\slashed{\partial}^\top\right)\zeta_2
+\left(Q_0+\slashed{x}\slashed{\partial}^\top-4\right)\zeta_3=0,
\end{equation}
\begin{equation}\label{upe3}
-2(1-2c)x.Z\zeta_1+cZ\cdot\partial^\top\zeta_1+(1-c)\left(Q_0+\slashed{x}\slashed{\partial}^\top\right)\zeta_2
+c\slashed{x}\left(4-\slashed{x}\slashed{\partial}^\top\right)\zeta_3=0.
\end{equation}
Equation (\ref{upe1}) can be rewritten as:
\begin{equation}\label{foeq}
\left(\slashed{x}\slashed{\partial}^\top-3\right)\left(-\slashed{x}\slashed{\partial}^\top+1\right)\zeta_1=0,
\end{equation}
so, there are two possibilities for the first-order field equations:
(1) the conformally coupled spinor field
\begin{equation} \label{zeta1equcon}
\left(\slashed{x}\slashed{\partial}^\top-1\right)\zeta_{1c}=0,\;\;\;\;
\left(Q_0-2\right)\zeta_{1c}=0,
\end{equation}
and (2) the minimally coupled spinor field
\begin{equation}\label{zeta1equ}
\left(\slashed{x}\slashed{\partial}^\top-3\right)\zeta_{1m}=0,\;\;
\;\; Q_0\zeta_{1m}=0.
\end{equation}
Their corresponding solutions are:
\begin{equation*}
\zeta_{1c}=\left(2-\slashed{x}\slashed{\partial}^\top\right){\cal U}
\phi_c,\;\;\; \zeta_{1m}=\slashed{x}\slashed{\partial}^\top{\cal U}
\phi_m=\slashed{x}\slashed{\partial}^\top{\cal U}
\left[Z\cdot\partial^\top+2Z\cdot x\right]\phi_c.
\end{equation*}
If $\zeta_1$ is a minimally coupled spinor field and
$c=\frac{2}{3}$, the equations (\ref{upe2}) and (\ref{upe3}) have solution and the spinor fields $\zeta_2$ and $\zeta_3$ can be
written in terms of $\zeta_{1m}$ as [Appendix D]:
\begin{equation}
\zeta_3=-\frac{1}{2}\left[\frac{3}{2}\slashed{x}x\cdot
Z+\frac{1}{6}\slashed{x}Z\cdot\partial^\top+\slashed{Z}\right]\zeta_{1m},
\end{equation}
and
\begin{equation}
\zeta_2=\left[\left(\frac{1}{2}+3w\right)x\cdot
Z+wZ\cdot\partial^\top-\frac{1}{6}\slashed{x}\slashed{Z}\right]\zeta_{1m},
\end{equation}
where $w$ is a constant arbitrary parameter.

\subsection{The general solution}

In this subsection, we want to find a general solution without any
conditions. This solution can be written as:
\begin{equation}\label{gs}
\Psi_\alpha=Z_{\alpha}^\top\psi_1+\nabla_\alpha^\top\psi_2+\gamma_\alpha^\top\psi_3.
\end{equation}
Similar to the previous subsection, by replacing (\ref{gs}) in the
field equation (\ref{II.12}), one obtained:
\begin{equation}\label{gsp1}
\left(Q_0+\slashed{x}\slashed{\partial}^\top-3\right)\psi_1=0,
\end{equation}
\begin{equation}\label{gsp2}
\slashed{x}x\cdot
Z\psi_1-\slashed{x}\left(4-\slashed{x}\slashed{\partial}^\top\right)\psi_2+\left(Q_0+\slashed{x}\slashed{\partial}^\top-4\right)\psi_3=0,
\end{equation}
\begin{equation}\label{gsp3}
-2(1-2c)x\cdot
Z\psi_1+cZ\cdot\partial^\top\psi_1+(1-c)\left(Q_0+\slashed{x}\slashed{\partial}^\top\right)\psi_2
+c\slashed{x}\left(4-\slashed{x}\slashed{\partial}^\top\right)\psi_3=0.
\end{equation}
The field equation (\ref{gsp1}) is similar to the field equation of
the spinor field $\zeta_1$ and there are two first order field
equations for $\psi_1$ as equation (\ref{foeq}). By using the identities (\ref{identity1}), (\ref{identity2}) and (\ref{identity3}) [Appendix D], the spinor fields $\psi_2$ and $\psi_3$ can be written in terms of
the spinor field $\psi_1$ as:
\begin{equation}
\psi_2=\left(n_1x\cdot Z +n_2Z\cdot
\partial^\top+n_3\slashed{x}\slashed{Z}\right)\psi_1,
\end{equation}
\begin{equation}
\psi_3=\left(m_1\slashed{x}x\cdot Z +m_2\slashed{x}Z\cdot
\partial^\top+m_3\slashed{Z}\right)\psi_1,
\end{equation}
where $n_1, ...,  m_3$ are the constant arbitrary parameters.

Replacing these solutions in equations (\ref{gsp2}) and
(\ref{gsp3}) and using the spinor field equation (\ref{zeta1equ}) for $\psi_1$, we obtain a solution only for the value
$c=\frac{2}{3}$:
\begin{equation*}
n_1=\frac{3}{4}+3t, \;\; n_2=\frac{1}{12}+t,
\;\;\;n_3=-\frac{1}{2}-4t,
\end{equation*}
\begin{equation*}
m_1=-\frac{5}{4}-6t, \;\; m_2=t, \;\;\;m_3=-\frac{3}{4}-3t,
\end{equation*}
where $t$ is a constant arbitrary parameter. In this case, the
general solution becomes:
\begin{equation*}
\Psi_\alpha=\left[Z_{\alpha}^\top+\nabla_\alpha^\top\left((\frac{3}{4}+3t)x\cdot
Z +(\frac{1}{12}+t)Z\cdot
\partial^\top-(\frac{1}{2}+4t)\slashed{x}\slashed{Z}\right)\right.\;\;\;\;\;\;\;\;\;\;\;\;\;\;\;\;\;\;\;
\end{equation*}
\begin{equation}\label{gsf}
\;\;\;\;\;\;\;\;\;\;\;\;\;\;\left.+\gamma_\alpha^\top\left(-(\frac{5}{4}+6t)\slashed{x}x\cdot
Z +t\slashed{x}Z\cdot
\partial^\top-(\frac{3}{4}+3t)\slashed{Z}\right)
\right]\psi_{1m}(x)\equiv\textbf{D}_\alpha(x,Z,t)\psi_{1m},
\end{equation}
where the spinor field $\psi_{1m}$ is
\begin{equation*}
\psi_{1m}=\slashed{x}\slashed{\partial}^\top{\cal U}
\phi_m=\slashed{x}\slashed{\partial}^\top{\cal
U}\left[Z'\cdot\partial^\top+2Z'\cdot x\right]\phi_c.
\end{equation*}
For this solution, there are a constant parameter $t$, a constant
spinor ${\cal U}$ and two constant five-vectors $Z^\alpha$ and
$Z'^\alpha$. One of the problem of this solution is that these constant parameters can not be fixed in the null curvature limit.
Another problem is that $c$ is fixed with value $\frac{2}{3}$, then the
solution is not a general solution and should be ignored.

Similar to the above procedure and using the spinor field equation (\ref{zeta1equcon}) for $\psi_1$, the general solution
becomes:
\begin{equation*}
\Psi_\alpha=\left[Z_{\alpha}^\top+\nabla_\alpha^\top\left\{n_1x\cdot
Z +n_2Z\cdot
\partial^\top+n_3\slashed{x}\slashed{Z} \right\}\right.\;\;\;\;\;\;\;\;\;\;\;\;\;\;\;\;\;\;\;\;\;\;\;\;\;\;\;\;\;\;\;\;\;
\end{equation*}
\begin{equation}\label{gsf}
\;\;\;\;\;\;\;\;\;\;\;\;\;\;\left.+\gamma_\alpha^\top\left\{m_1\slashed{x}x\cdot
Z+m_2\slashed{x}Z\cdot
\partial^\top+m_3\slashed{Z}\right\} \right]\psi_{1c}(x)\equiv\textbf{D}_\alpha(x,Z,c)\psi_{1c},
\end{equation}
where
\begin{equation*}
n_1=\frac{c(4c-1)-1}{2(1-c)}, \;\; n_2=\frac{3c-2}{4(1-c)},
\;\;\;n_3=-\frac{c(4c-3)}{4(1-c)},
\end{equation*}
\begin{equation*}
m_1=-\frac{c^2(2c+3)+2(1-3c)}{c(1-c)}, \;\;
m_2=-\frac{c(16c-23)+8}{4(1-c)}, \;\;\;m_3=\frac{4c(1-c)-1}{4(1-c)}.
\end{equation*}
The spinor field $\psi_{1c}$ is:
\begin{equation*}
\psi_{1c}=\left(2-\slashed{x}\slashed{\partial}^\top\right){\cal U}
\phi_c.
\end{equation*}
In this case, there are a constant spinor ${\cal U}$ and a constant
five-vector $Z_\alpha$, which can  be fixed in the null curvature
limit and specify the indecomposable representation of de Sitter
group. There exists a five-dimensional trivial representation with
respect to $Z_\alpha^{(\lambda)}$ \cite{gaha}. For a thorough
investigation regarding the five existing polarization states $\lambda =
0,1, 2, 3, 4$, the reader may refer to \cite{gaha}. Solution
(\ref{gsf}) is also a general solution since $c$ is arbitrary. In
the next section, the quantum field operator and its corresponding
two-point function are constructed by this solution.

\section{Quantum field operator and Two-point function}

In the previous section, it is proved that the ambient space
formalism permits us to write the vector-spinor field in terms of a
vector-spinor polarization state and a massless conformally coupled
scalar field (\ref{gsf}):
\begin{equation}\label{gspcs}
\Psi_\alpha\equiv {\cal D}_\alpha (x,\partial,Z){\cal U}\phi_c,
\end{equation}
where
\begin{equation*}
{\cal
D}_\alpha(x,\partial,Z)=\textbf{D}_\alpha(x,Z,c)(2-\slashed{x}\slashed{\partial}^\top).
\end{equation*}
First, we recall the construction of the quantum field operator and
the two-point function for the massless conformally coupled scalar
field $\phi_c$, then it is simply generalized to the massless vector
-spinor field.

\subsection{Massless conformally coupled scalar field}

In ambient space formalism the massless conformally coupled scalar field solution can be written in terms of de Sitter plane wave $(x\cdot \xi)^{\sigma}$ with $\sigma=-1,-2$ \cite{Bros 1}. This plane waves solution can not be defined globally in de Sitter space, but it can be defined
globally in complex de Sitter space-time \cite{Bros 1,Bros 2, Bros
3}:
\begin{equation*}
M_H^{(c)}=\left\{ z=x+iy\in \C^5;\;\;\eta_{\alpha \beta}z^\alpha
z^\beta=(z^0)^2-\vec z\cdot\vec
z-(z^4)^2=-H^{-2}\right\}
\end{equation*}
\begin{equation}
=\left\{ (x,y)\in \R^5\times \R^5;\;\; x^2-y^2=-H^{-2},\; x\cdot
y=0\right\}.
\end{equation}
Let $T^\pm= \R^5+iV^\pm$ to be the forward and backward tubes in $
\C^5$. The domain $V^+$(resp. $V^-)$ stems from the causal structure
on $M_H$:
\begin{equation}\label{v+-}
V^\pm=\left\{ x\in \R^5;\;\; x^0\stackrel{>}{<} \sqrt {\parallel
\vec x\parallel^2+(x^4)^2}\right\}.
\end{equation}
Then we introduce their respective intersections with $M_H^{(c)}$
\begin{equation}
{\cal T}^\pm=T^\pm\cap M_H^{(c)},
\end{equation}
which are called the forward and backward tubes of the complex de
Sitter space $M_H^{(c)}$. Finally, the ``tuboid'' on
$M_H^{(c)}\times M_H^{(c)}$ is defined as:
\begin{equation}\label{tuboid}
{\cal T}_{12}=\left\{ (z,z');\;\; z\in {\cal T}^+,z' \in {\cal T}^-
\right\}.
\end{equation}
If $z$ varies in ${\cal T}^+$
(or ${\cal T}^-$) and $\xi$ lies in the positive cone ${\cal C}^+$ (\ref{poscon}):
\begin{equation*}
\xi \in {\cal C}^+=\left\{ \xi \in {\cal C}; \; \xi^0>0
\right\},
\end{equation*}
the plane wave solutions are globally defined since the imaginary
part of $(z.\xi)$ has  a fixed sign (for more details, see
\cite{Bros 2}). In terms of de Sitter complex plane wave, the field
operator can be written in the following form \cite{Takook
77,taro15}:
\begin{equation} \label{csfoinco}
\phi_c(z)=\sqrt{ c_0 }\int_{S^3} d\mu({\bf \xi}) \left\lbrace\;
a({\bf\tilde{\xi}})(z\cdot\xi)^{-2} +a^\dag({\bf
\xi})(z\cdot\xi)^{-1} \right\rbrace,
\end{equation}
where $\xi^\alpha=(1, \vec \xi, \xi^4)$, $\tilde \xi^\alpha=(1,
-\vec \xi, \xi^4)$ and the vacuum state is defined as \cite{Takook
77}:
\begin{equation*}
a({\bf \xi})|\Omega>=0,\;\; a^\dag({\bf \xi})|\Omega>=|\xi>, \;\;<
\xi' |\xi>=\delta_{S^3}(\xi-\xi'), \;\;\int_{S^3}   d\mu({\bf
\xi})\delta_{S^3}(\xi-\xi') =1.
\end{equation*}
The notations are defined explicitly in \cite{Takook 77}.

The analytic two-point function is defined in terms of the complex
de Sitter plane waves by \cite{Bros 1,Bros 2}:
\begin{equation}\label{tpfscinint}
W_c(z_1,z_2)=\left<\Omega|\phi(z_1)\phi(z_2)|\Omega\right>=c_0\int_{S^3}d\mu(\xi)
(z_1\cdot\xi)^{-2}(z_2.\cdot\xi)^{-1},
\end{equation}
and $c_0$ is obtained by using the local Hadamard condition. The
vacuum state $|\Omega>$ in this case is exactly equivalent to the
Bunch-Davies vacuum state \cite{Takook 77}. One can easily calculate
(\ref{tpfscinint}) in terms of the generalized Legendre function
\cite{Bros 2}:
\begin{equation}\label{atpfc}
W_c(z_1,z_2)=\frac{-iH^2}{2^4\pi^2} P_{-1}^{(5)}(H^2 z_1\cdot z_2)=
\frac{H^2}{8\pi^2}\frac{-1}{1-{\cal
Z}(z_1,z_2)}=\frac{H^2}{4\pi^2}(z_1-z_2)^{-2},
\end{equation}
where ${\cal Z}(z_1,z_2)=-H^2 z_1\cdot z_2$. The Wightman two-point
function ${\cal W}_c(x_1,x_2)$ is the boundary value (in the sense
of its interpretation as a distribution function, according to the
theorem A.2 in \cite{Bros 2}) of the function $W_c(z_1, z_2)$ which
is analytic in the domain ${\cal T}_{12}$ of $M_H^{(c)}\times
M_H^{(c)}$ \cite{Bros 2}. The boundary value is defined for $z_1
=x_1+iy_1\in {\cal T}^-$ and $z_2=x_2+iy_2\in {\cal T}^+$ as
\begin{equation*}
{\cal Z}(z_1,z_2)={\cal Z}(x_1,x_2)-i\tau\epsilon(x_1^0,x_2^0),
\end{equation*}
where $y_1=(-\tau,0,0,0,0)\in V^-$, $y_2=(\tau,0,0,0,0)\in V^+$ and
$\tau \rightarrow 0$. Then, one obtains \cite{Bros 2,Takook
th,ChTa}:
\begin{equation*}
{\cal W}_c(x_1,x_2)=\frac{-H^2}{8\pi^2}\lim_{\tau \rightarrow
0}\frac{1}{1-{\cal
Z}(x_1,x_2)+i\tau\epsilon(x_1^0,x_2^0)}
\end{equation*}
\begin{equation}\label{stpci2}
=\frac{-H^2}{8\pi^2}\left[ P\frac{1}{1-{\cal Z}(x_1,x_2)}
-i\pi\epsilon(x_1^0,x_2^0)\delta(1-{\cal
Z}(x_1,x_2))\right],
\end{equation}
where the symbol $P$ is the principal part and ${\cal Z}(x_1,x_2)$
is the geodesic distance between two points $x_1$ and $x_2$ on the de
Sitter hyperboloid:
\begin{equation*}
{\cal Z}(x_1,x_2)=-H^2 x_1\cdot
x_2=1+\frac{H^2}{2} (x_1-x_2)^2,
\end{equation*}
and
\begin{equation}
\epsilon(x_1^0-x_2^0)=\left\{\begin{array}{clcr} 1&x_1^0>x_2^0
\\
0&x_1^0=x_2^0\\  -1&x_1^0<x_2^0\\    \end{array} \right..
\end{equation}

\subsection{Massless Vector-spinor field}

Using the equations (\ref{gspcs}) and (\ref{csfoinco}), in the complex de Sitter space, the vector-spinor field operator is
then defined as \cite{Takook 77}:
\begin{equation}
\Psi_\alpha (z)\equiv\sqrt{ c_0 } \int_{S^3}
d\mu(\xi)\sum_{\lambda=0}^4\sum_{r=1,2}{\cal D}_\alpha (z,\partial,Z^\lambda) {\cal U}^r(\xi)
\left\lbrace\; a(\tilde{\xi})(z\cdot\xi)^{-2}
+a^\dag(\xi)(z\cdot\xi)^{-1}
\right\rbrace,
\end{equation}
where the explicit form of ${\cal U}^r$ is defined in \cite{Takook th,sep}. The explicit form of the polarization five-vector $Z^\lambda$
depends on the indecomposable representation of de Sitter group \cite{gaha}. As
a simple case, one can chose \cite{Takook 77,taro15}:
\begin{equation}\label{zpolar}
\sum_{\lambda=0}^4 \sum_{\lambda'=0}^4 Z^{(\lambda)}_\alpha
Z^{(\lambda')}_\beta=\eta_{\alpha\beta},\;\;\; Z^{(\lambda)}\cdot
Z^{(\lambda')}=\eta ^{\lambda\lambda'}.
\end{equation}
The analytic function $S_{\alpha\beta}(z,z')$ is defined as \cite{sep}
\begin{equation*}
S_{\alpha\beta}(z,z')=\langle\Omega|\Psi_\alpha(z)\bar\Psi_\beta(z')|\Omega\rangle,
\end{equation*}
where $z,z'\in M_H^c$ and $|\Omega\rangle$ is the vacuum state. By
using the identity \cite{Takook th}
\begin{equation*}
\sum_r {\cal U}^r\otimes\bar{\cal U}^r=\slashed{\xi}
\gamma^4,
\end{equation*}
the two-point function can be written in the following compact form:
\begin{equation} \label{tpfs3-2}
S_{\alpha\beta}(z,z')=\sum_{\lambda,\lambda'}{\cal D}_\alpha(z,\partial,Z^\lambda)
S_c(z,z'){\bar{\cal D}_\beta}(z',\overleftarrow \partial',Z^{\lambda'}),
\end{equation}
where $\bar{\cal D}=\gamma^0\gamma^4{\cal D}^\dag \gamma^0\gamma^4$
and $S_{c}$ is the two-point function of massless conformally spinor
field \cite{Takook th}:
\begin{equation}
S_{c}({z},{z'})=\left(-\slashed{z'}\slashed{\partial'}^\top+1
\right)\gamma^4W_c(z,z').
\end{equation}
The two point function of massless conformally coupled scalar field ($W_c)$ and massless spinor field ($S_c$) are constructed on Bunch-Davis vacuum states and preserve the Hadamard structure \cite{Bros 1,Bros 2, sep}. Then our two-point function (\ref{tpfs3-2}), which  is constructed from a massless conformally coupled scalar field and a polarization tensor-spinor, preserve the correct Hadamard structure. since the polarization tensor-spinor take the derivative of the $W_c$ and the derivative can not break the Hadamard structure. It is important to note that in our construction the negative norm states does not appear for the scalar field (\ref{csfoinco}) and the Bunch-Davis vacuum states is used.

\section{Conclusions}

In this paper, the massless vector-spinor or super-gauge field
$\Psi_\alpha$ is studied in de Sitter ambient space formalism. The
super-gauge invariant Lagrangian density is presented by using the
super-gauge covariant derivative. The Gupta-Bleuler triplet is
discussed. It is shown that the field solutions are built up from a
conformally coupled scalar field and a vector-spinor polarization
state. Finally, the  quantum field operator and its corresponding
two-point function are calculated. The two-point function is
analytic in this construction. Since the quantum field theory in our
formalism is unitary and analytic, a unitary supergravity in de
Sitter ambient space formalism seems quite plausible. 

In this paper the free field quantization is considered, using the interaction Lagrangian which is defined in ambient space notation by the gauge principle \cite{Takook 77} one can perform the one-loop correction for various fields that (may) couple to the gravitino. By coupling
this vector-spinor gauge field with the massless spin-$2$ gauge
field in de Sitter ambient space formalism, a unitary supergravity
may be obtained, which will be studied in a forthcoming paper. 


\vspace{0.5cm} \noindent {\bf{Acknowlegements}}: We are grateful
to  S. Teymourpoor for her interest in this work.
\appendix
\section{The Euler-Lagrange equation}

From the action (\ref{action}), the Lagrangian density is
\begin{equation}\label{A.1}
{\cal L}=\left(\tilde{\nabla}^\top_\alpha\tilde{
\Psi}_\beta-\tilde{\nabla}^\top_\beta\tilde{
\Psi}_\alpha\right)\left( \nabla^{\top\alpha}
\Psi^\beta-\nabla^{\top\beta} \Psi^\alpha\right),
\end{equation}
where
\begin{equation*}
\left(\tilde{\nabla}^\top_\alpha\tilde{
\Psi}_\beta-\tilde{\nabla}^\top_\beta\tilde{ \Psi}_\alpha\right)
=\left(\partial_{\alpha}^{\top}\tilde{ \Psi}_\beta-x_\beta\tilde{
\Psi}_\alpha-\partial_{\beta}^{\top}\tilde{
\Psi}_\alpha+x_\alpha\tilde{ \Psi}_\beta\right).
\end{equation*}
Using the Euler-Lagrange equation
\begin{equation*}
\frac{\delta {\cal L}}{\delta\tilde{
\Psi}_m}-\partial_{l}^{\top}\frac{\delta {\cal
L}}{\delta(\partial^{\top}_{ l}\tilde{ \Psi}_m)}=0,
\end{equation*}
we obtain
\begin{equation*}
\frac{\delta {\cal L}}{\delta\tilde{ \Psi}_
{m}}=(x_\alpha\delta_{\beta}^{m}-x_\beta\delta_{\alpha}^{m}) \left(
\nabla^{\top\alpha} \Psi^\beta-\nabla^{\top\beta}
\Psi^\alpha\right),
\end{equation*}
and
\begin{equation*}
\frac{\delta {\cal L}}{\delta(\partial^{\top}_{ l}\tilde{ \Psi}_{m})}=
(\delta_{\alpha}^{l}\delta_{\beta}^{m}-\delta_{\beta}^{l}\delta_{\alpha}^{m})\left(
\nabla^{\top\alpha} \Psi^\beta-\nabla^{\top\beta}
\Psi^\alpha\right).
\end{equation*}
Then the  Euler-Lagrange equation leads immediately to the
following field equation
\begin{equation*}
(x_\alpha-\partial_\alpha^{\top}) \left(\nabla^{\top\alpha}
\Psi^\beta-\nabla^{\top\beta }\Psi^\alpha \right)=0,
\end{equation*}
which is equation (\ref{vsfe0}).

\section{gauge invariant}

The Lagrangian density
\begin{equation*}
{\cal L}=\left(\tilde{\nabla}^\top_\alpha\tilde{
\Psi}_\beta-\tilde{\nabla}^\top_\beta\tilde{
\Psi}_\alpha\right)\left( \nabla^{\top\alpha}
\Psi^\beta-\nabla^{\top\beta} \Psi^\alpha\right),
\end{equation*}
is invariant under the following gauge transformations:
\begin{equation}\label{B.1}
\Psi_\alpha \longrightarrow
\Psi^g_\alpha=\Psi_\alpha+\nabla^\top_{\alpha } \psi,
\end{equation}
\begin{equation}\label{B.2}
\tilde{\Psi}_\alpha \longrightarrow
\tilde{\Psi}^g_\alpha=\tilde{\Psi}_\alpha+\partial_{\alpha}^{\top}\tilde{\psi}.
\end{equation}
The Lagrangian density can be divided up into two parts
\begin{equation}\label{B.3}
A=\left( \nabla^{\top\alpha} \Psi^\beta-\nabla^{\top\beta}
\Psi^\alpha\right)=\left(\partial^{\top\alpha}
\Psi^\beta+\gamma^\alpha\slashed{x}\Psi^\beta -\partial^{\top\beta}
\Psi^\alpha-\gamma^\beta\slashed{x}\Psi^\alpha \right),
\end{equation}
\begin{equation}\label{B.4}
B=\left(\tilde{\nabla}^\top_\alpha\tilde{
\Psi}_\beta-\tilde{\nabla}^\top_\beta\tilde{ \Psi}_\alpha\right)
=\left(\partial_{\alpha}^{\top}\tilde{ \Psi}_\beta-x_\beta\tilde{
\Psi}_\alpha-\partial_{\beta}^{\top}\tilde{
\Psi}_\alpha+x_\alpha\tilde{ \Psi}_\beta\right),
\end{equation}
where any parts have their own gauge transformation. Under the gauge
transformation \eqref{B.1}, the first part \eqref{B.3} becomes:
\begin{equation*}
A^g=\partial^{\top\alpha}( \Psi^\beta+\nabla^{\top{\beta }}
\psi)+\gamma^\alpha\slashed{x}( \Psi^\beta+\nabla^{\top{\beta }}
\psi) -\partial^{\top\beta} ( \Psi^\alpha+\nabla^{\top{\alpha }}
\psi)-\gamma^\beta\slashed{x}( \Psi^\alpha+\nabla^{\top{\alpha }
}\psi)
\end{equation*}
\begin{equation*}
=\partial^{\top\alpha} \Psi^\beta+\gamma^\alpha\slashed{x}\Psi^\beta
-\partial^{\top\beta}
\Psi^\alpha-\gamma^\beta\slashed{x}\Psi^\alpha+\partial^{\top\alpha}\nabla^{\top{\beta
} }\psi+\gamma^\alpha\slashed{x}\nabla^{\top{\beta }}
\psi-\partial^{\top\beta}\nabla^{\top{\alpha }
}\psi-\gamma^\beta\slashed{x}\nabla^{\top{\alpha }} \psi.
\end{equation*}
Using the following relations
\begin{equation*}
\partial^{\top\alpha}\nabla^{\top{\beta }}
\psi=\partial^{\top\alpha}\partial^{\top\beta}\psi+\gamma^\beta\gamma^\alpha\psi
+\gamma^{\beta}x^{\alpha}\slashed{x}\psi+\gamma^{\beta}\slashed{x}\partial^{\top\alpha}\psi-\eta^{\alpha\beta}\psi-x^{\alpha}x^{\beta}\psi-x^{\beta}\partial^{\top\alpha}\psi,
\end{equation*}
\begin{equation*}
\partial^{\top\beta}\nabla^{\top{\alpha }}
\psi=\partial^{\top\beta}\partial^{\top\alpha}\psi+\gamma^\alpha\gamma^\beta\psi+\gamma^{\alpha}x^{\beta}\slashed{x}\psi+\gamma^{\alpha}\slashed{x}\partial^{\top\beta}\psi-\eta^{\beta\alpha}\psi-x^{\beta}x^{\alpha}\psi-x^{\alpha}\partial^{\top\beta}\psi,
\end{equation*}
\begin{equation*}
\gamma^\alpha\slashed{x}\nabla^{\top{\beta }}
\psi=\gamma^\alpha\slashed{x}\partial^{\top\beta}\psi+\gamma^\alpha\slashed{x}
x^{\beta}\psi+\gamma^\alpha\gamma^\beta\psi,
\end{equation*}
\begin{equation*}
\gamma^\beta\slashed{x}\nabla^{\top{\alpha }
}\psi=\gamma^\beta\slashed{x}\partial^{\top\alpha}\psi+\gamma^\beta\slashed{x}
x^{\alpha}\psi+\gamma^\beta\gamma^\alpha\psi,
\end{equation*}
one can obtain the identity:
\begin{equation}\label{B.5}
\partial^{\top\alpha}\nabla^{\top{\beta } }\psi+\gamma^\alpha\slashed{x}\nabla^{\top{\beta }} \psi-\partial^{\top\beta}\nabla^{\top{\alpha
}} \psi-\gamma^\beta\slashed{x}\nabla^{\top{\alpha }} \psi=0.
\end{equation}
Therefore, by using \eqref{B.5}, we can see that \eqref{B.3} is
invariant under \eqref{B.1}. Similarly for \eqref{B.4}, under the transformation \eqref{B.2}, we have
\begin{equation*}
B^g=\partial_{\alpha}^{\top}(\tilde{\Psi}_\beta+\partial_{\beta}^{\top}\tilde{\psi})
-x_\beta(\tilde{\Psi}_\alpha+\partial_{\alpha}^{\top}\tilde{\psi})
-\partial_{\beta}^{\top}(\tilde{\Psi}_\alpha+\partial_{\alpha}^{\top}\tilde{\psi})
+x_\alpha(\tilde{\Psi}_\beta+\partial_{\beta}^{\top}\tilde{\psi})
\end{equation*}
\begin{equation*}
=\partial_{\alpha}^{\top}\tilde{ \Psi}_\beta-x_\beta\tilde{
\Psi}_\alpha-\partial_{\beta}^{\top}\tilde{
\Psi}_\alpha+x_\alpha\tilde{
\Psi}_\beta+\partial_{\alpha}^{\top}\partial_{\beta}^{\top}\tilde{\psi}-
x_\beta\partial_{\alpha}^{\top}\tilde{\psi}-\partial_{\beta}^{\top}\partial_{\alpha}^{\top}\tilde{\psi}
+x_\alpha\partial_{\beta}^{\top}\tilde{\psi}.
\end{equation*}
Using the identity
\begin{equation*}
[\partial_{\alpha}^{\top},\partial_{\beta}^{\top}]=x_\beta\partial_{\alpha}^{\top}-x_\alpha\partial_{\beta}^{\top},
\end{equation*}
 or equivalently
\begin{equation*}
\partial_{\alpha}^{\top}\partial_{\beta}^{\top}\tilde{\psi}
-x_\beta\partial_{\alpha}^{\top}\tilde{\psi}-\partial_{\beta}^{\top}\partial_{\alpha}^{\top}\tilde{\psi}
+x_\alpha\partial_{\beta}^{\top}\tilde{\psi}=0,
\end{equation*}
one can see that \eqref{B.4} is also invariant under the gauge transformation (\ref{B.2}).


\section{Divergence spinor state}
The divergence of the field equation \eqref{II.12} and using the definition $Q_{\frac{3}{2}}^{(1)}$, we obtain:
\begin{equation} \label{C.1}
{\partial^{\top\alpha}}\left[Q_0\Psi_\alpha^d+\slashed{x}\slashed{\partial}^\top\Psi_\alpha^d-\frac{11}{2}\Psi_\alpha^d+2x_\alpha
\partial^\top\cdot\Psi^d+\gamma_\alpha\slashed{\Psi}^d\right]+\frac{5}{2}{\partial^{\top}}\cdot\Psi^d
+c{\partial^{\top\alpha}}(\nabla_{\alpha}^\top
\partial^\top\cdot\Psi^d)=0.
\end{equation}
By the supplementary identities
\begin{equation*}
{\partial^{\top\alpha}}(Q_0\Psi_\alpha^d)=\left(Q_0-6\right)\partial^\top\cdot\Psi^d,
\end{equation*}
\begin{equation*}
{\partial^{\top\alpha}}(\slashed{x}\slashed{\partial}^\top\Psi_\alpha^d)
=\left(1+\slashed{x}\slashed{\partial}^\top\right)\partial^\top\cdot\Psi^d-\slashed{\partial}^\top\slashed{\Psi}^d,
\end{equation*}
\begin{equation*}
\partial^{\top\alpha}\left(\gamma_\alpha\slashed{x}\partial^\top\cdot\Psi^d\right)
=\left(4-\slashed{x}\slashed{\partial}^\top\right)\partial^\top\cdot\Psi^d,
\end{equation*}
one can write \eqref{C.1} in the form
\begin{equation*}
(1-c)\left(Q_0+\slashed{x}\slashed{\partial}^\top\right)\partial^\top\cdot\Psi^d=0,
\end{equation*}
or equivalently as:
\begin{equation*}
(1-c)\left(Q_{\frac{1}{2}}^{(1)}+\frac{5}{2}\right)\partial^\top\cdot\Psi^d=0.
\end{equation*}

\section{$\Psi^d_\alpha$ with $c=\frac{2}{3}$}

The condition $\slashed{\Psi}^{d}= 0$ on equation (\ref{dps}) permit
us to obtain a relation between the three spinor fields $\zeta_1$, $\zeta_2$ and $\zeta_3$:
\begin{equation}\label{zeta123}
\left(4-\slashed{x}\slashed{\partial}^\top\right)\zeta_2=
\slashed{x}\slashed{Z}\zeta_1-x\cdot Z\zeta_1+4\slashed{x}\zeta_3,
\end{equation}
and by using equations (\ref{zeta123}) and (\ref{upe2}), $\zeta_3$
satisfies:
\begin{equation}\label{zeta2}
\left(Q_0+\slashed{x}\slashed{\partial}^\top\right)\zeta_3
=\slashed{x}\slashed{\partial}^\top\left(4-\slashed{x}\slashed{\partial}^\top\right)\zeta_3=-\left(\slashed{Z}+2\slashed{x}x\cdot
Z\right)\zeta_1.
\end{equation}
Now we should invert (\ref{zeta2}) to determine $\zeta_3$ in terms
of $\zeta_1$. At the first stage, one can rewrite (\ref{zeta2}) as
follows:
\begin{equation}
\left(4-\slashed{x}\slashed{\partial}^\top\right)\zeta_3=-{\left(\slashed{x}\slashed{\partial}^\top\right)}^{-1}\left(\slashed{Z}+2\slashed{x}x\cdot
Z\right)\zeta_1.
\end{equation}
If we define the field equation (\ref{foeq}) as
$\slashed{x}\slashed{\partial}^\top\zeta_1=a\zeta_1$ with $a=1,3$,
one can prove the following identities:
\begin{equation}\label{identity1}
\slashed{x}\slashed{\partial}^\top\slashed{x}x\cdot Z \zeta_1=
\left((5-a)\slashed{x} x\cdot Z +\slashed{Z}\right)\zeta_1,
\end{equation}
\begin{equation}\label{identity2}
\slashed{x}\slashed{\partial}^\top\slashed{x} Z\cdot
\partial^\top \zeta_1= \left(-2a\slashed{x}x\cdot Z +(3-a)\slashed{x}Z
\cdot
\partial^\top-a\slashed{Z}\right)\zeta_1,
\end{equation}
\begin{equation}\label{identity3}
\slashed{x}\slashed{\partial}^\top\slashed{Z}\zeta_1=\left(2a\slashed{x}x\cdot
Z +2\slashed{x}Z\cdot\partial^\top+a\slashed{Z}\right)\zeta_1.
\end{equation}

By using the above identities, one can find that there exists a
solution only for $a=3$ as (the minimally coupled spinor field):
\begin{equation}\label{solzeta32}
\zeta_3=-\frac{(3n+5)}{4}\slashed{x}x\cdot Z\zeta_{1m}
-\frac{(n+1)}{4}\slashed{x}Z\cdot
\partial^\top\zeta_{1m}-\frac{1}{2}\slashed{Z}\zeta_{1m},
\end{equation}
where $n$ is an arbitrary constant parameter.

Now we determine $\zeta_2$ in terms of $\zeta_{1m}$. Putting
(\ref{solzeta32}) into (\ref{upe3}) leads to
\begin{equation}\label{zeta22}
\left(Q_0+\slashed{x}\slashed{\partial}^\top\right)\zeta_2=\frac{2-c(3n+5)}{1-c}x.Z\zeta_{1m}-\frac{c(1+n)}{1-c}
Z\cdot\partial^\top\zeta_{1m}.
\end{equation}
Identities \eqref{identity1}-\eqref{identity3} for $a=3$ can be
written equivalently as follows:
\begin{equation*}
\slashed{x}\slashed{\partial}^\top x\cdot Z \zeta_{1m}=
\left(2x\cdot Z +\slashed{x}\slashed{Z}\right)\zeta_{1m},
\end{equation*}
\begin{equation*}
\slashed{x}\slashed{\partial}^\top Z\cdot \partial^\top \zeta_{1m}=
\left(6x\cdot Z
+4Z\cdot\partial^\top-3\slashed{x}\slashed{Z}\right)\zeta_{1m},
\end{equation*}
\begin{equation*}
\slashed{x}\slashed{\partial}^\top\slashed{x}\slashed{Z}\zeta_{1m}=\left(6x\cdot
Z +2Z\cdot\partial^\top+\slashed{x}\slashed{Z}\right)\zeta_{1m}.
\end{equation*}
The first order field equation for the spinor field $\zeta_2$ is
obtained as:
\begin{equation}\label{zeta2eq2}
\left(4-\slashed{x}\slashed{\partial}^\top\right)\zeta_2=\frac{10-c(3n+13)}{4(1-c)}x.Z\zeta_{1m}+\frac{2-c(n+3)}{4(1-c)}Z\cdot\partial^\top\zeta_{1m}
-\slashed{x}\slashed{Z}\zeta_{1m}.
\end{equation}
This equation has a solution only for the values $c=\frac{2}{3}$ and
$n=-\frac{2}{3}$:
\begin{equation}\label{zeta2sol}
\zeta_2=\left((\frac{1}{2}+3w)x\cdot
Z+wZ\cdot\partial^\top-\frac{1}{6}\slashed{x}\slashed{Z}\right)\zeta_{1m},
\end{equation}
where $w$ is another arbitrary constant parameter.


\end{document}